# First Demonstration of Robust Tri-Gate $\beta$-Ga$_2$O$_3$ Nano-membrane Field-Effect Transistors Operated Up to 400 °C

Hagyoul Bae[1], Tae Joon Park[2], Jinhyun Noh[1], Wonil Chung[1],
Mengwei Si[1], Shriram Ramanathan[2], and Peide D. Ye[1,*]

[1]School of Electrical and Computer Engineering and Birck Nanotechnology Center, [2]School of Materials Engineering,
Purdue University, West Lafayette, IN 47907 USA
Email: yep@purdue.edu

*Abstract*—Nano-membrane tri-gate $\beta$-gallium oxide ($\beta$-Ga$_2$O$_3$) field-effect transistors (FETs) on SiO$_2$/Si substrate fabricated via exfoliation have been demonstrated for the first time. By employing electron beam lithography, the minimum-sized features can be defined with a 50 nm fin structure. For high-quality interface between $\beta$-Ga$_2$O$_3$ and gate dielectric, atomic layer-deposited 15-nm-thick aluminum oxide (Al$_2$O$_3$) was utilized with Tri-methyl-aluminum (TMA) self-cleaning surface treatment. The fabricated devices demonstrate extremely low subthreshold slope (*SS*) of 61 mV/dec, high drain current ($I_{DS}$) ON/OFF ratio of $1.5 \times 10^9$, and negligible transfer characteristic hysteresis. We also experimentally demonstrated robustness of these devices with current–voltage (*I–V*) characteristics measured at temperatures up to 400 °C.

## I. Introduction

$\beta$-Ga$_2$O$_3$ is one of the promising materials for next-generation power electronics owing to its ultra-wide bandgap of 4.6–4.9 eV, high breakdown electric field of 8 MV/cm, high electron mobility of 100–150 cm$^2$/V·s, and sustainability for high-temperature operation and mass production with low-cost fabrication [1]-[8]. In addition, the $\beta$-Ga$_2$O$_3$ material has a higher Baliga's figure-of-merit (FOM) than that of silicon carbide (SiC) and gallium nitride (GaN) [9], [10]. Owing to these advantages, monolithic β-Ga$_2$O$_3$ transistors could also be considered for high-temperature operation in harsh environments. In particular, stable operation of electronic devices in severe conditions, mainly at high temperatures, is indispensable for many applications in the defense, automotive, nuclear instrumentation, and aerospace fields [11], [12]. Recently, several studies have demonstrated improved electrical performances of $\beta$-Ga$_2$O$_3$ FETs by using double-gate [13], multi-channel with wrap-gate [14], vertical channel [15], back-gate [16], and recessed-gate [17], [18] devices. In particular, among these advanced technologies, structure innovation to enhance gate controllability to boost higher current density and suppress interface or short-channel effect becomes critical during device research [19]-[20]. There are still opportunities to achieve better switching characteristics, higher integration density, and lower power consumption in $\beta$-Ga$_2$O$_3$ materials and device development.

In this study, the fabrication and performance of tri-gate $\beta$-Ga$_2$O$_3$ FETs with single and multi-fin channel structures formed from nano-membranes are presented. In the single fin-structure, a narrow channel with a fin width ($W_{fin}$) of 50 nm is fabricated to exploit the high ON/OFF ratio of $I_{DS}$ and superior subthreshold slope, while maintaining reliable performance at temperature from room temperature (RT) to 400 °C.

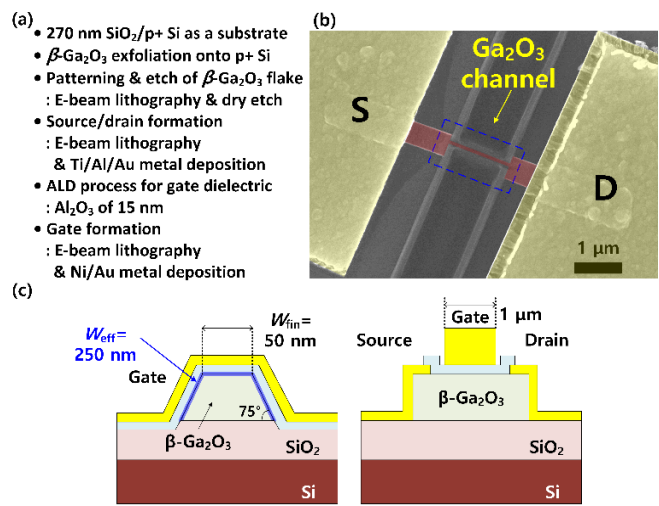

Fig. 1. (a) Process flow for device fabrication of exfoliated tri-gate $\beta$-Ga$_2$O$_3$ nano-membrane FETs with the top-gate structure. (b) SEM image of a fabricated device. (c) Cross-sectional schematics along both channel width and length directions.

## II. Device Fabrication

Fig. 1(a) lists the key fabrication steps for the tri-gate nano-membrane $\beta$-Ga$_2$O$_3$ FETs on the SiO$_2$/Si substrate. For the fabrication of the top-gate devices, thin (100) $\beta$-Ga$_2$O$_3$ nano-membranes with a Sn doping concentration of $2.7 \times 10^{18}$ cm$^{-3}$ was transferred from the bulk $\beta$-Ga$_2$O$_3$ substrate onto a p+ Si wafer with a 270 nm SiO$_2$ as gate dielectric. The fin-shape active channel region was defined by electron (e)-beam lithography and the dry-etch process. To form narrower $\beta$-Ga$_2$O$_3$ fin structure, we used a BCl$_3$/Ar gas mixture in an inductively coupled plasma-reactive ion etching (ICP-RIE) system (Panasonic E620 Etcher) for 15 min [21]. The etching rate of (100) β-Ga₂O₃ is about 10 nm/min under process conditions: RF power of 100 W; BCl₃ flow of 15 sccm; Ar flow of 60 sccm; and pressure of 0.6 Pa. Subsequently, the source (S) and drain (D) regions were formed via e-beam lithography patterning, Ti/Al/Au (15 /60 /50 nm) metallization, and lift-off process. No post-deposition thermal annealing was performed. By employing the atomic layer deposition (ALD) process, high-quality Al₂O₃ gate dielectric was deposited to minimize gate leakage current and high-quality interface by self-cleaning effect using TMA as the precursor [22], [23]. Subsequently, e-beam lithography was carried out for gate patterning and Ni/Au (50 /80 nm) metal gate was deposited via an e-beam evaporator. Fig. 1(b) presents the scanning electron microscope (SEM) image showing the etched β-Ga₂O₃ channel with $W_{fin}$ of 50 nm, a gate length ($L_G$) of 1 μm, and a fin height ($H_{fin}$) of 95 nm. As the fabricated device has a 3D fin structure, the total effective channel width ($W_{eff}$) is approximately 250 nm, which is 5 times wider than the footprint $W_{fin}$ of 50 nm. Fig. 1(c) shows the cross-sectional view of the fabricated device along both channel width and length directions. Using atomic force microscopy (AFM), the measured physical thickness of the exfoliated β-Ga₂O₃ nano-membrane is approximately 95 nm.

## III. EXPERIMENTAL RESULTS AND DISCUSSION

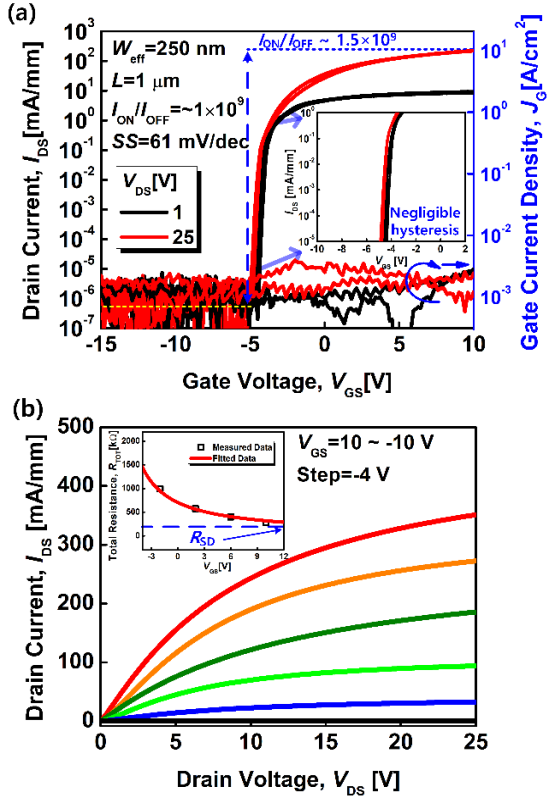

Fig. 2. (a) Measured $I_{DS}$–$V_{GS}$ transfer characteristics and (b) $I_{DS}$–$V_{DS}$ output characteristics of the fabricated tri-gate β-Ga₂O₃ FET.

Fig. 2 (a) shows the measured transfer characteristics ($I_{DS}$–$V_{GS}$) of the fabricated tri-gate β-Ga₂O₃ FET with $W_{eff}$ of 250 nm and $L_G$ of 1 μm, showing the following excellent electrical performances: 1) $I_{ON}$=350 mA/mm normalized with the $W_{eff}$; 2) $I_{ON}/I_{OFF}$=1.5×10⁹; 3) $SS_{min}$=61 mV/dec; 4) DIBL=12 mV/V; 5) $V_{hys}$=30 mV. The electrical characterizations were performed using a Keysight B1500 semiconductor parameter analyzer, a Keithley 4200-SCS parameter analyzer with a high-temperature measurement system (Micromanipulator H1000 Thermal Chuck System), and a Cascade Summit probe station. Fig. 2 (b) shows the measured output characteristics ($I_{DS}$-$V_{DS}$) as a function of $V_{GS}$ from 0 to –20 V. A maximum drain current density ($I_{DS\_max}$) of 350 mA/mm in the fabricated device on the SiO₂/Si substrate is obtained, which is higher than that obtained in our previous study of top-gate devices on SiO₂/Si substrate [24]. This proposed fin structure enables to control β-Ga₂O₃ channel from three sides of the gate and improve the gate electrostatics significantly as in Si CMOS technology. The threshold voltage ($V_T$) is determined to be -2.8 V by the constant current method. The parasitic source/drain resistances ($R_{SD}$) is obtained by extrapolating $V_{GS}$ based on channel resistance method (CRM) as shown in the inset of Fig. 2 (b) [25]. The $R_C$ and sheet resistance ($R_{SH}$) is extracted to be 8.0 Ω·mm and 6.2 kΩ/□, respectively. Further studies on β-Ga₂O₃ contacts are highly demanded in the development of β-Ga₂O₃ device technology [26].

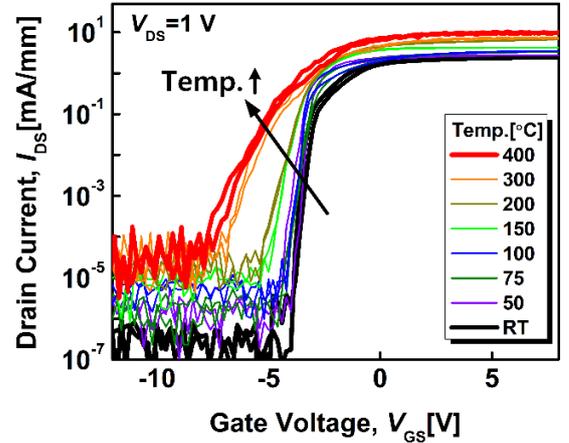

Fig. 3. Measured $I_{DS}$-$V_{GS}$ characteristics of the fabricated tri-gate β-Ga₂O₃ FETs on SiO₂/Si substrate at various temperatures ranging from RT to 400 °C.

The measured $I_{DS}$–$V_{GS}$ characteristics of the fabricated tri-gate β-Ga₂O₃ FET on the SiO₂/Si substrate measured at various temperatures ranging from RT to 400 °C are shown in Fig. 3 with negligible hysteresis. Although $I_{OFF}$ starts to gradually increase as the temperature increases, our proposed tri-gate β-Ga₂O₃ FETs have more stable characteristics for temperatures up to 400 °C compared to previous results [6], [13]. The variations in the extracted SS, $V_T$, field-effect mobility ($μ_{FE}$), and $I_{ON}/I_{OFF}$, as a function of temperature are plotted in Fig. 4. The value of SS increases from 61 mV/dec to 710 mV/dec and $V_T$ shifts toward a negative direction due to thermally excited carriers from interface states between the channel and the gate dielectric, as shown in Fig. 4 (a) and (b). The interface trap density ($D_{it}$ [eV$^{-1}$cm$^{-2}$]) of the fabricated devices is extracted to $1.0 \times 10^{11}$ eV$^{-1}$cm$^{-2}$ at RT. The significant increase of SS beyond Boltzmann thermal limit indicates a large amount $D_{it}$ of $1.6 \times 10^{13}$ eV$^{-1}$cm$^{-2}$ could be activated at 400 °C. In addition, $\mu_{FE}$ decreases also due to the increased $D_{it}$ at the interface and the phonon scattering in channel at high temperatures as shown in Fig. 4 (c) [6], [13], [27]. We should also note that, even at 400 °C, $I_{ON}/I_{OFF}$ was observed to be approximately $3 \times 10^5$, as shown in Fig. 4 (d). The tri-gate $\beta$-Ga$_2$O$_3$ FETs with ALD Al$_2$O$_3$ as gate dielectric demonstrate robust electrical performances at high temperatures. Furthermore, the measured I-V characteristics at RT were again obtained after cooling the device from high temperature to RT.

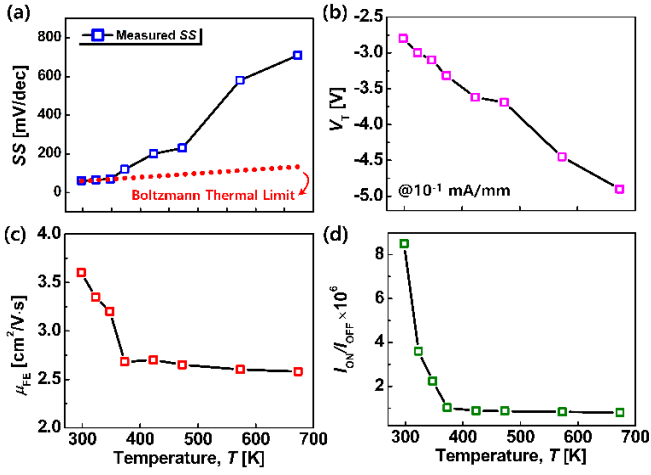

Fig. 4. Temperature dependences (RT–400 °C) at $V_{DS}$ = 1 V of typical device parameters such as (a) SS (dashed line: Boltzmann thermal limit of SS), (b) $V_T$, (c) $\mu_{FE}$, and (d) $I_{ON}/I_{OFF}$ of the tri-gate $\beta$-Ga$_2$O$_3$ FETs on the SiO$_2$/Si substrate.

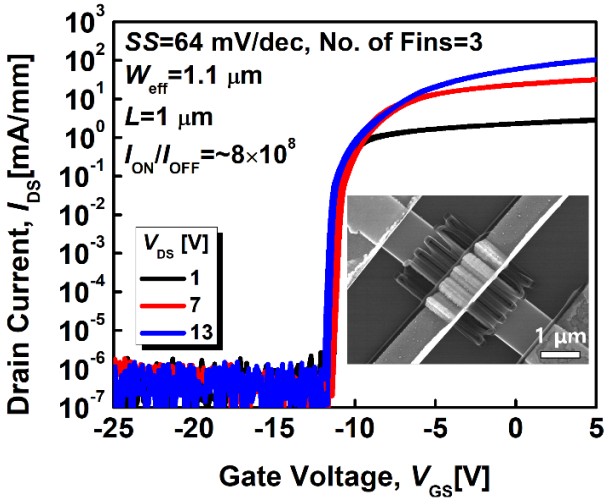

Fig. 5. Measured $I_{DS}$-$V_{GS}$ characteristics of the fabricated $\beta$-Ga$_2$O$_3$ FETs with multi-fins. Inset shows SEM image for top view of the fabricated device with 3 fins.

## IV. EXPERIMENTAL RESULTS AND DISCUSSION

As shown in Fig. 5, we also demonstrated the feasibility of the multi-fin tri-gate $\beta$-Ga$_2$O$_3$ FETs for a high integration density. The inset shows the SEM image for top view of the fabricated device with 3 fin channels. In case of multi-fin devices, overall device performances are also comparable to the single-fin devices. To provide clear evidences for advantages of tri-gate $\beta$-Ga$_2$O$_3$ FETs, we fabricated three types of devices with different gate structures ((i) tri-gate, (ii) planar-gate, and (iii) bottom-gate) under same process conditions and investigated the impact of the tri-gate $\beta$-Ga$_2$O$_3$ FETs via comparison with other devices, as shown in Fig. 6. Table 1 shows the comparison data of the fabricated devices with different gate structures. It is noteworthy that the tri-gate $\beta$-Ga$_2$O$_3$ FETs have high channel ratio ($W_{eff}$/perimeter of channel width ($W_{peri}$)) of 0.85 and aspect ratio (AR) of 2 resulting in better electrical performances such as $I_{ON}/I_{OFF}$, SS, and $D_{it}$ [28], [29]. In this regard, minimizing the device degradation caused by interface states and modulating the effective charges from $\beta$-Ga$_2$O$_3$ channel in tri-gate $\beta$-Ga$_2$O$_3$ FETs can be important since the switching characteristics are very strongly influenced by interface between $\beta$-Ga$_2$O$_3$ and bottom SiO$_2$ substrate.

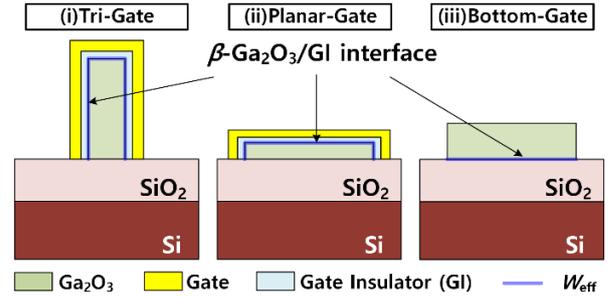

Fig. 6. Schematics of cross-sectional view for the fabricated devices with different gate structures: (i) tri-gate, (ii) planar-gate, (iii) bottom-gate.

TABLE 1
Comparison Data with Different Gate Structures

| @$V_{DS}$=1 V | Sample #1 | Sample #2 | Sample #3 |
|---|---|---|---|
| Structure | Single Fin | Planar | Bottom-Gate |
| Gate Dielectric | Al$_2$O$_3$ | Al$_2$O$_3$ | SiO$_2$ |
| $I_{ON}/I_{OFF}$ | ~1.3×10$^7$ | ~9.5×10$^6$ | ~1.2×10$^6$ |
| $SS_{min}$ [mV/dec] | 61 | 100 | 250 |
| Type | Flake | Flake | Flake |
| $D_{it}$ [eV$^{-1}$cm$^{-2}$] | ~1×10$^{11}$ | ~9×10$^{11}$ | ~1×10$^{12}$ |
| $W_{eff}$ | 250 nm | 1 µm | 1.5 µm |
| Channel Ratio ($W_{eff}/W_{peri}$) | 0.85 | 0.55 | 0.45 |
| Aspect Ratio (AR) | 2 | 0.12 | 0.1 |

A benchmark of the fabricated tri-gate $\beta$-Ga$_2$O$_3$ FETs is summarized in Table II. The overall electrical performances of the fabricated tri-gate $\beta$-Ga$_2$O$_3$ FETs with a single channel are improved over different types of devices reported previously.

In particular, our proposed device is highly scalable with an active channel area of 0.05 μm$^2$ by employing an extremely scaled β-Ga$_2$O$_3$ fin structure. The $D_{it}$ of the fabricated device shows a high-quality interface compared with previously reported results [30], [31].

TABLE 2
Benchmarking for Electrical Performances of Ga$_2$O$_3$ FETs

|  | This Work | Ref. [13] | Ref. [14] | Ref. [15] |
|---|---|---|---|---|
| $I_{ON}/I_{OFF}$ (Max.) | ~1.5×10$^9$ | ~7×10$^7$ | ~5×10$^5$ | ~1×10$^9$ |
| $SS_{min}$ [mV/dec] | 61 | 70 | 158 | 200 |
| # of Fin | Single | - | 48 | 20 |
| Type | Flake | Flake | MOVPE | Epitaxial Growth |
| $D_{it}$ (Max.) [eV$^{-1}$cm$^{-2}$] | ~1×10$^{11}$ | - | ~1×10$^{12}$ | - |
| Temp. | RT~400 °C | RT~250 °C | RT | RT |
| $V_{hys}$ | 30 mV | 30 mV | 700 mV | 100 mV |
| $I_{DS\_max}$ | 350 mA/mm | 1 mA/mm | 3 mA/mm | 1 kA/cm$^2$ |
| $W_{eff}$ | 250 nm | 7 μm | 24 μm | 50 μm |

## V. CONCLUSION

In this study, top-gate tri-gate nano-membrane β-Ga$_2$O$_3$ FETs were successfully demonstrated for the first time. Outstanding electrical performances were achieved from the single fin-like structure with a $W_{fin}$ of 50 nm. The fabricated devices have a superior subthreshold slope and $I_{ON}/I_{OFF}$ resulting from enhanced top gate controllability. Moreover, the fabricated devices demonstrate sustainable reliability under high temperatures of up to 400 °C, validating its use in applications involving harsh environments. Multi-fin devices also represent improved electrical performances comparable to single-fin devices for high integration density. Consequently, we expect that tri-gate β-Ga$_2$O$_3$ FETs have the potential as a low-cost and high-performance power device technology after the establishment of epitaxy materials.